\documentclass[showpacs,preprint,amsmath,amssymb]{revtex4}

  \usepackage[dvips]{graphics}

\usepackage{graphicx}

\begin{document}

\title{Laser-controlled local magnetic field with semiconductor
quantum rings}

\author{Yuriy V. Pershin}
\email{pershin@pa.msu.edu}
\author{Carlo Piermarocchi}

\affiliation{Department of Physics and Astronomy, Michigan State
University, East Lansing, Michigan 48824}

\begin{abstract}
We analize theoretically the dynamics of $N$ electrons localized in a
semiconductor quantum ring under a train of phase-locked infrared
laser pulses.  The pulse sequence is designed to control the total
angular momentum of the electrons. The quantum ring can be put in
states characterized by strong currents.
The local magnetic field created by these currents can be used for a
selective quantum control of single spins in semiconductor systems.
\end{abstract}

\date{\today}

\maketitle

The quantum control by trains of phase-locked laser pulses is a
powerful and intuitive technique that has been applied to many atomic
and molecular systems. It has been successfully employed in
controlling the angular momentum of Rydberg electrons and molecular
wavepackets~\cite{verlet02}, and it was the key technique in the
creation of Schr\"odiger cat's states in Rydberg
atoms~\cite{noel96}. Semiconductor quantum dots and rings are
artificial atoms with energy levels that can be engineered, and the
realization of optical control in these systems is particularly
appealing for quantum device applications.

\begin{figure}[bp]
\centering
\includegraphics[height=11.0cm,angle=270]{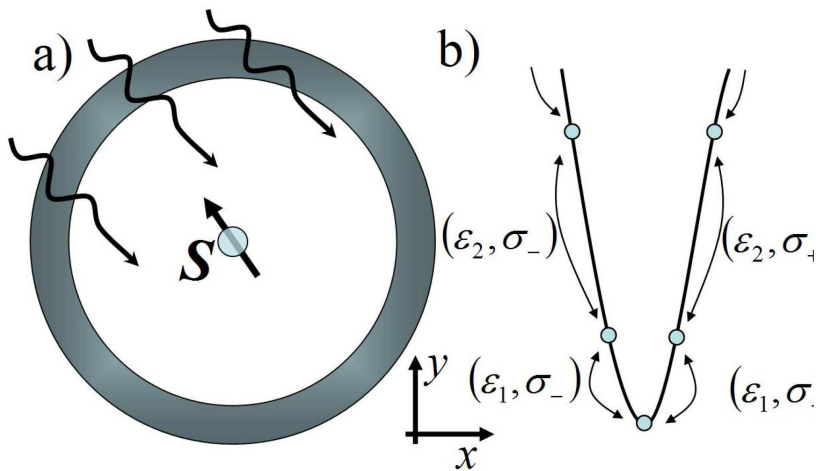}
\caption{(a) Scheme of the system: circularly polarized light
control a spin by exciting currents in a quantum ring. (b) Scheme
of the energy levels and selection rules for N=1.}
\label{fig1}
\end{figure}

In this paper we propose a scheme involving phase-locked infrared
pulses to control the total angular momentum of N electrons in a
quantum ring. This implies that a strong current can be generated in
the ring. The latter can be exploited to generate and control a local
magnetic field in spin-based quantum
computers~\cite{kane98,privman98}.  A possible application of the
radiation-induced currents to single spin control is shown in
Fig.~\ref{fig1}(a). A spin, provided e.g. by a magnetic impurity, is
embedded at the center or on top of a narrow quantum ring and can be
locally controlled by the magnetic field due to the current in the
ring. A scheme that uses arrays of parallel wires to create localized
magnetic fields has been recently proposed~\cite{Lidar}. The
significant advantage of our laser-controlled approach is that it does
not require external leads: the magnetic field is controlled by laser
pulses. The peculiar orbital properties of the many-body eigenstates
in a quantum ring and Pauli blocking effects make this scheme robust
against the relaxation of the current by phonon and photon emission.

Small semiconductor nanorings can be made by the same self-assembly
methods used for the fabrication of quantum dots. Persistent currents
due to the Aharonov-Bohm flux have been observed in these quantum
structures in the presence of an external magnetic flux along the ring
axis \cite{lorke00}. Currently, there is great theoretical interest in
persistent
currents~\cite{VRWZ98,Qian,Serega,Manolescu,PCwithSO,Peeters,PCquantc,ulloa97}. In
particular, it is believed that they can be generated, besides using an
external magnetic flux, also by the interplay of spin-orbit interaction
and hyperfine coupling~\cite{VRWZ98,Qian}. We propose here that
circularly polarized radiation can create currents in a quantum ring
without an external magnetic flux.  The circular polarization of the
light propagating along the axis of the ring breakes the
clockwise-anticlockwise symmetry of the many-electron wavefunction. In
the case of a cw excitation these currents are persistent currents:
they are associated to the ground state of the ring {\it dressed} by
the external laser field. Pulsed lasers can however produce stronger
currents in a shorter time, which is more interesting for our
purpose. We will discuss the cw case elsewhere. Notice that linearly
polarized excitation will not induce a current but can affect the
current induced by an external magnetic flux~\cite{Manolescu}.

We start from $\Psi_0(\varphi_1,\cdots \varphi_N)$, the exact
N-electron ground state of a narrow ring Hamiltonian
\begin{equation}
H_R=-\frac{\hbar^2}{2m^*a^2} \sum\limits_{i=1}^N
\frac{\partial^2}{\partial \varphi^2_i} +U(\varphi_1, \cdots,
\varphi_N)~. \label{HL}
\end{equation}
$a$ is the ring radius, $m^*$ is the electron effective mass and
$\varphi_i$ are the axial coordinates of the
electrons. $U(\{\varphi_i\})$ is the electron-electron Coulomb
interaction. The operator $\hat{L}_z=-i \sum_{i=1}^{N}
\partial/\partial \varphi_i$ commutes with the $H_R$ and $\hat{L}_z
\Psi_0=0$. We want to design a multi-pulse control Hamiltonian of the
form
\begin{equation}
H_C(t)= \sum_j \mathbf{d}\cdot\mathbf{E}_j(t-\tau_j)~,
\end{equation}
where $ \mathbf{d}$ is the total dipole moment of the $N$ electrons
and $\mathbf{E}_j$ is the electric field of the $j$-th pulse.  The
target states in our scheme are written as
\begin{equation}
\Psi_l= e^{i l (\varphi_1+\cdots+\varphi_N)} \Psi_0~,
\label{compact}
\end{equation}
with $l$ integer, and are exact eigenstates of the Hamiltonian in
Eq.~(\ref{HL}) with energy $E_l=E_0+\hbar^2 N l^2/(2 m^* a^2)$ and
with $\hat{L}_z \Psi_l=N l \Psi_l$.  These states are called {\it
compact} states and they can be seen as rigid rotational modes of
the many-body electron system as a whole. The wave functions
$\Psi_0$ and $\Psi_l$ are linked by a rotating wave transformation
and their Coulomb correlation properties are identical.
Every time a $\sigma_+$ polarized photon is
absorbed or emitted the total $L_z$ increases or decreases by one,
while the opposite is valid for $\sigma_-$.
\begin{figure}[tbp]
\centering
\includegraphics[height=11cm,angle=270]{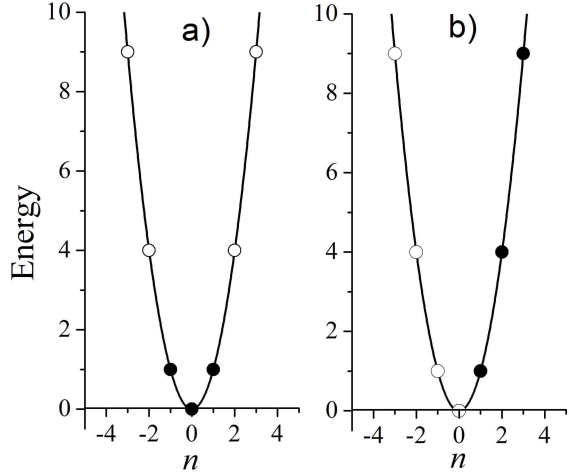}
\caption{Energy level population scheme: the ground state (a) and
a compact excited state (b). The equilibrium ground state (a) is characterized
by zero current. The current in the excited state (b) is strong since
all electrons occupy the states with the same sign of the angular
momentum.} \label{fig2}
\end{figure}

Let us consider first the dynamics of a single electron confined in
the ring neglecting  relaxation processes. The energy, wave
function, and current of an electron in the $n$-th level are given by
\begin{equation}
\varepsilon_{n}=\frac{\hbar^2n^2}{2m^*a^2},\quad
\psi_{n}=\frac{1}{\sqrt{2\pi a}}e^{in\varphi} , \quad
i_n=\frac{e\hbar}{2\pi a^2 m^*}n. \label{energ}
\end{equation}
The time-dependent electric field $\mathbf{E}_j(t)$ generates
transitions between the levels $\varepsilon_n$. We write the electric
field associated to the $j$-th pulse as $\mathbf{E}_j(t)=E_{0j}
\cos(\omega_j t)\hat{x}+E_{1j} \sin(\omega_j t)\hat{y}$ which accounts
for arbitrary polarization. $E_1=E_0$ corresponds to $\sigma_+$
polarization and $E_1=-E_0$ to $\sigma_-$ polarization. Matrix
elements for dipole transitions are non-zero only between nearest
single-particle states $\langle n|-d E_\pm | n+1 \rangle=eaE_0e^{\pm i
\omega t}/2$, where $\pm$ denotes $\sigma_\pm$ radiation. By writing
the electron wave function in the form $\Psi=\sum c_n(t) \psi_n$,
where $c_n^2(t)$ gives the probability to find the electron in the
state $n$, we get, for $\sigma_+$ polarization, the equation of motion
\begin{figure}[tbp]
\centering \includegraphics[height=11cm,angle=270]{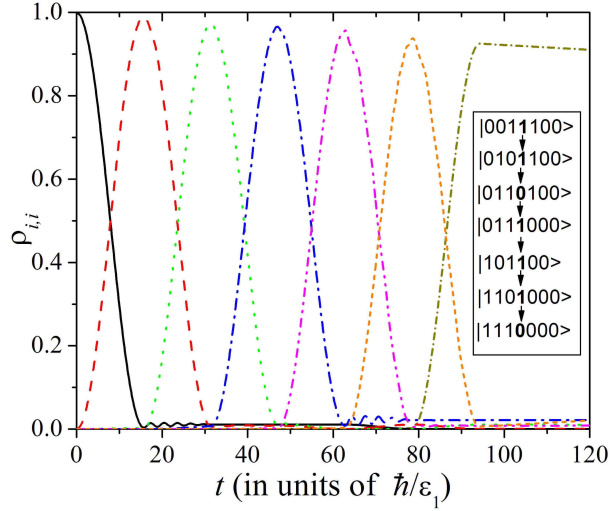}
\caption{(Color online) Evolution of the population of 3-electron
levels induced by the pulse sequence. Inset: ideal sequence of
transitions. The Rabi energy in the calculation is
$\alpha=0.2\varepsilon_1$, and the relaxation rate is
$\gamma=0.02\varepsilon_1$.} \label{fig3}
\end{figure}
\begin{equation}
i\hbar\dot{c}_n=\varepsilon_n c_n+\frac{\alpha}{2} e^{i\omega t}c_{n+1}+\frac{\alpha}{2}
e^{-i\omega t}c_{n-1}~, \label{ci}
\end{equation}
where $\alpha=eaE_0$ is the Rabi energy. The unitary transformation
$c_n=e^{-in\omega t}b_n$ allows us to eliminate the exponential
factors in Eq.~(\ref{ci})
\begin{equation}
i\hbar\dot{b}_n=\left(\varepsilon_n-n\hbar\omega \right)b_n+\frac{\alpha}{2}
b_{n+1}+\frac{\alpha}{2} b_{n-1}. \label{bi}
\end{equation}
From Eq. (\ref{bi}) it follows that resonant transitions between
two levels occur when $\tilde \varepsilon_n=\tilde
\varepsilon_{n+1}$, where $\tilde
\varepsilon_n=\varepsilon_n-n\hbar\omega$. Notice that $\tilde
\varepsilon_{n+1}-\tilde \varepsilon_{n}<0$ for any $n<0$ and
$\omega>0$, meaning that $\sigma_+$ radiation can only excite resonant
transitions between levels with $n\geq 0$. For $\sigma_-$ polarization,
the dynamics of $b_n$ is described by Eq. (\ref{bi}) with $\omega$
replaced by $-\omega$. It can be shown that
$\sigma_-$ radiation can only excite resonanant transitions between
levels with $n\leq 0$.  The selection rules for single electron transitions are shown in
Fig. \ref{fig1}(b).

Consider now $N$-electron states of spinless and non-interacting
electrons.
Starting from a ground state, like the one in Fig. \ref{fig2}(a)
for $N$=3, and using a sequence of pulses, our goal is to put the
electrons in a compact state that is characterized by a strong
current like e.g. the one in Fig. \ref{fig2}(b). If in the ground
state the levels $n=-N_1$, ..., $N_1$ are populated, in the final
compact state the levels $n=(-N_1+p)$, ..., $(N_1+p)$ are
populated, i.e. $p$ electrons are moved from the states with
negative angular momentum to the states with positive angular
momentum. This is accomplished using a sequence of $\pi$ pulses
with frequency and polarization $(\omega_{N+1},\sigma_+)$,
$(\omega_{N},\sigma_+)$, ..., $ (\omega_1,\sigma_+)$, $
(\omega_1,\sigma_-)$, ..., $(\omega_N,\sigma_-)$,
$(\omega_{N+2},\sigma_+)$, ..., where $\hbar
\omega_n=\varepsilon_n-\varepsilon_{n-1}$.  Notice that (i) each
pulse increases $L_z$ by one by stimulated absorption of a
$\sigma_+$ or stimulated emission of a $\sigma_-$ photon, and (ii)
at each step in the sequence the energy of the optical transition
is the {\it smallest} allowed by the many particle energy
configuration on condition that $L_z$ increases by one. This
implies that in our path toward the target compact states we
follow the lowest energy states with fixed total $L_z$ ({\it
yrast} line of excitations). From an operative standpoint, the
polarization and the energy of the $j$ pulse in the sequence can
be directly determined by spectroscopic measurements on the
system. The energy of the photon to be used corresponds to
$min[\omega^+_<,\omega^-_<]$ where $\omega^+_<$ is the lowest
$\sigma_+$-polarized peak in absorption, and $\omega^-_<$ is the
lowest $\sigma_-$-polarized peak in emission/gain. In the ideal
case the final current is
\begin{equation}
I=\sum\limits_{n=-N_1+p}^{N_1+p}j_n=p(2N+1)i_1 \label{icurrent}~,
\end{equation}
which is $\simeq 2p$ times stronger than the typical amplitude of the
persistent current oscillations induced by the Aharonov-Bohm flux.

The general idea of optical transitions with the smallest exchange
of energy can be generalized to include the spin of the electrons.
Notice that   in the absence of spin-orbit coupling the total spin
of the electrons is conserved in the transitions. We have found
that if a spin-orbit coupling of the Rashba form is present, this
control scheme can create spin-polarized currents in the
ring~\cite{pershinfuture}. Many-body calculations done using the
configuration interaction method~\cite{viefers04} show that the
yrast line of excitations is always well separated from high
excited states in small rings. Therefore in that case our general
prescription for the pulse sequence to target compact states will
not be affected by the presence of Coulomb interaction. The total
exchange of energy at the end of the pulse sequence does not
depend on the Coulomb interaction. The electromagnetic field
provides only the kinetic energy necessary to a rigid rotation of
all the electrons, since the potential energy of all compact
states is the same of the one in the ground state.

Off-resonant population transfer and relaxation will affect our
control scheme. We give a full description of the electron dynamics in
the ring using a density matrix formalism. We include all non-resonant
transitions without rotation wave approximation in the simulation. The
equation of motion of the density matrix is given by
$\dot\rho=-i/\hbar\left[ H,\rho\right]+D \left\{ \rho \right\}$ where
the relaxation superoperator $D \left\{ \rho \right\}$ has the form
$
D \left\{ \rho \right\}= \sum\limits_\xi \left( \mathbf{L}_\xi \rho
\mathbf{L}^\dag_\xi-\frac{1}{2}\mathbf{L}^\dag_\xi \mathbf{L}_\xi \rho
-\frac{1}{2}\rho \mathbf{L}^\dag_\xi \mathbf{L}_\xi \right)~,$
with Lindblad operators $\mathbf{L}_\xi=\gamma_{\xi}|m\rangle\langle
n|$ describing decoherence and relaxation processes~\cite{lindblad76}.
\begin{figure}[tbp]
\centering
\includegraphics[height=11cm,angle=270]{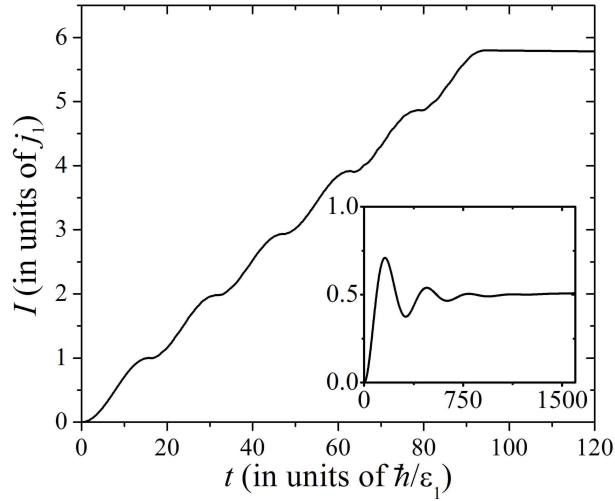}
\caption{Current in the ring as a function of time using the proposed
pulse sequence. The inset show for comparison the current in the case
of a cw excitation.}
\label{fig4}
\end{figure}

In order to demonstrate our approach, we solve the Liouville
equation numerically assuming that 3 electrons are in the ring.
Limiting our consideration to $7$ possible single-particle states
(shown in Fig. \ref{fig2}), the ground state of the system can be
represented as $|001\mathbf{1}100\rangle$, where the $1$ indicates
an occupied level and $0$ an empty level.  The $n=0$ level is
identified by a bold-face index.  There are 35 possible states
that can be obtained with different occupations of 7 levels by 3
electrons. We start from the configuration where only the ground
state is populated, assuming that $\varepsilon_2-\varepsilon_1 \gg
k_B T$. The optical control is realized with pulses of rectangular
shape. We select the pulse duration $T_p$ to obtain $\pi$ pulses
as $\alpha T_p/ \hbar=\pi$.

Fig. \ref{fig3} shows the result of our calculations. The applied
pulse sequence generates almost perfect transitions between the
states. The diagonal matrix elements in Fig. \ref{fig3} correspond
to the states indicated in the inset.
We note that off-resonance transitions and relaxation result in
the decrease of the maximum of population inversion at long times.
The small fast oscillations of the diagonal density matrix
elements (clearly resolved, for example, for
$|001\mathbf{1}100\rangle$ at $t\sim 20$) are due to
off-resonanant transitions. The current in the ring is calculated
as $I=\textnormal{tr}\left( \rho j\right)$, where $j$ is the
quantum mechanical current operator, and is shown in Fig.
\ref{fig4}. Its saturation value at $t \gtrsim 100$ is close to
the ideal value $I=6i_1$ predicted by Eq. (\ref{icurrent}). We
show for comparison in the inset of Fig.~\ref{fig4} the current
obtained with a cw excitation. The initial oscillation is due to
the Rabi oscillation between two states, but the equilibrium value
gives a finite value for the current. The Rabi energy in the inset
is $\alpha=0.02 \varepsilon_1$.

The magnetic field along the axis of the ring generated by a
current $I$ in the ring can be estimated using $
\mathbf{B}=\mathbf{z}\mu_0Ia^2/\left(2\left(
z^2+a^2\right)^{\frac{3}{2}}\right)$ where $\mu_0$ is the magnetic
constant and $\mathbf{z}$ is the unit vector along the axis of the
ring. For a GaAs-based quantum rings of $a=10$nm ($m^*=0.067m_e$)
one finds $i_1=0.44\mu$A. The magnetic field at the center of the
ring, taking into account two-fold spin degeneracy and assuming
$N_1=5$ and $p=5$, gives $B(z=0)\approx 3$mT. This value is of the
same order of magnitude of local magnetic field obtained by arrays
of nanowires~\cite{Lidar}. The total switching time $T_{sw}$ is
given by $T_{sw}=p(2N_1+1)T_p$. Our numerical simulations show
that the total current decreases if the Rabi energy becomes
comparable to the transition energy.  In our simulation we have
used $\alpha=0.2\varepsilon_1$ which gives $T_p \approx 1.8$ps and
$T_s \approx 100$ps.

The photon spontaneous emission rate is $\gamma^2_{\xi}\propto d^2
\omega^3_\xi$, with $\hbar \omega_\xi=\varepsilon_m-\varepsilon_n$,
and $d$ the dipole moment between the states $|m\rangle$ and
$|n\rangle$. Given a typical radius of 10 nm, and $\hbar \omega_\xi$
of the order of 5 meV, the spontaneous emission rate can be estimated
in the order of hundreds of microseconds. It is expected that the main
relaxation mechanism is provided by the emission of acoustic phonons
through the deformation potential interaction. We have estimated the
rate of a single electron relaxation by single phonon emission and
found a characteristic relaxation time of the order of 1 to 10
ns. Notice that the the phonon needs to provide angular momentum to
the electron system and this decreases the matrix element between the
states $|m\rangle$ and $|n\rangle$ with increase of $|n-m|$. This
property in combination with the Pauli exclusion principle allows to
conclude that the compact state in Fig.~ 2~(b) is robust against
relaxation by phonon emission in the sense that the relaxation time of
such a state is much longer than the typical single-electron
relaxation time. In the simulation above we have used a conservative
value for the relaxation rate of $\gamma=0.025 \varepsilon_1$,
corresponding to about $500 ps$, which is shorter than the estimated
values for the photon and phonon assisted relaxation times. For larger
rings the energy transitions involved in the present scheme will be in
the microwave region of the electromagnetic spectrum. Several
experiments of quantum control in that spectral range have been
reported~\cite{mani,Zudov}.



In conclusion, we have shown that a strong current can be excited
in a quantum ring via a train of phase-locked infrared pulses. The
key component in our scheme is the circularly polarization of
pulses, which increases the angular momentum of the many-electron
state in the ring. We propose to use the system as
externally-controlled source of local magnetic fields for
single-spin quantum logic.

This research was supported by the National Science Foundation,
Grant NSF DMR-0312491.

\end{document}